\documentstyle[12pt]{article}
\newcommand{\be}{\begin{equation}}
\newcommand{\ee}{\end{equation}}
\newcommand{\bea}{\begin{eqnarray}}
\newcommand{\eea}{\end{eqnarray}}
\newcommand{\bt}{\begin{tabular}}
\newcommand{\et}{\end{tabular}}
\newcommand{\ba}{\begin{array}}
\newcommand{\ea}{\end{array}}

\newcommand{\bvec}{\vec}

\newcommand{\uam}{u^{\mu}} 
\newcommand{\uan}{u^{\nu}}
\newcommand{\ubm}{u_{\mu}}
\newcommand{\ubn}{u_{\nu}}
\newcommand{\famn}{F^{\mu \nu}}
\newcommand{\fbmn}{F_{\mu \nu}}
\newcommand{\ftamn}{\widetilde{F}^{\mu \nu}}
\newcommand{\ftbmn}{\widetilde{F}_{\mu \nu}}
\newcommand{\eam}{E^{\mu}}
\newcommand{\ean}{E^{\nu}}
\newcommand{\ebm}{E_{\mu}}
\newcommand{\ebn}{E_{\nu}}
\newcommand{\bam}{B^{\mu}}
\newcommand{\ban}{B^{\nu}}
\newcommand{\bbm}{B_{\mu}}
\newcommand{\bbn}{B_{\nu}}
\newcommand{\pam}{\psi^{\mu}}
\newcommand{\pan}{\psi^{\nu}}
\newcommand{\pbm}{\psi_{\mu}}

\newcommand{\dbm}{\partial_{\mu}}
\newcommand{\dbn}{\partial_{\nu}}
\newcommand{\gam}{\gamma^{\mu}}
\newcommand{\gan}{\gamma^{\nu}}

\newcommand{\al}{\bvec{\alpha}}
\newcommand{\ta}{\bvec{\tau}}

\begin{document}
\setcounter{page}{0}
\thispagestyle{empty}
\baselineskip=20pt

\hfill{
\begin{tabular}{l}
DSF$-$97/14 \\
hep-th/9704144
\end{tabular}}

\bigskip\bigskip

\begin{center}
\begin{huge}
{\bf Covariant Majorana Formulation of Electrodynamics }
\end{huge}
\end{center}

\vspace{2cm}

\begin{center}
{\Large
Salvatore Esposito\\} 
\end{center}

\vspace{0.5truecm}

\normalsize
\begin{center}
{\it
Dipartimento di Scienze Fisiche, Universit\`a di Napoli,
Mostra d'Oltremare Pad. 19, 80125 Napoli Italy} \\
and\\
{\it INFN, Sezione di Napoli,
Mostra d'Oltremare Pad. 20, I-80125 Napoli Italy}\\
e-mail: sesposito@na.infn.it
\end{center}

\vspace{3truecm}

\begin{abstract}
We construct an explicit covariant Majorana formulation of Maxwell 
electromagnetism which does not make use of vector 4-potential. This 
allows to write a ``Dirac'' equation for the photon containing all the 
known properties of it. In particular, the spin and (intrinsic) boost 
matrices are derived and the helicity properties of the photon are 
studied.
\end{abstract}

\newpage

\section{Introduction.}

Recently, many different experiments have revealed new electromagnetic 
phenomena that, although they can be well described by Maxwell 
electromagnetism, sound quite unusual with respect to our traditional 
view of electrodynamics.

An example is the observation of tunneling photons with group 
velocities also greater than $c$; this effect does not violate 
Einstein causality principle and can be described in terms of Maxwell 
equations (see \cite{Espo, Mugnai} and references therein).

Another long discussed example is the experimental measurement of the 
Evans-Vigier $B^3$ field in non-linear optical experiments (such as in 
the inverse Faraday effect, etc.) \cite{Evans}, the $B^3$ field being 
the phase free spin field of Maxwell electromagnetism. To this regard, 
Evans has interestingly noted \cite{Priv} that the cyclic equations 
for the $B^3$ field \cite{Evans} are particularly evident in the 
Majorana formulation of electrodynamics.
\footnote{Although more indirectly, also in the theoretical 
description of tunneling photons the employment of the writing of 
Maxwell equations as a Dirac-like equation for the photon seems quite 
useful; this can be seen confronting \cite{Mugnai} and \cite{Kac}. The 
author is grateful to A. Ranfagni for bringing to his attention these 
two references.}

Majorana's original idea \cite{Majo} was that if the Maxwell theory of 
electromagnetism has to be viewed as the wave mechanics of the photon, 
then it must be possible to write Maxwell equations as a Dirac-like 
equation for a probability quantum wave $\psi$, this wave function 
being expressable by means of the physical ${\cal \bvec{E}}$, ${\cal 
\bvec{B}}$ fields. This can be, indeed, realized introducing the 
quantity
\be
\bvec{\psi} \; = \; {\cal \bvec{E}} \; - \; i \, {\cal \bvec{B}}
\label{11}
\ee
since $\bvec{\psi}^{\ast} \cdot \bvec{\psi} \, = \, {\cal \bvec{E}}^2 
\, + \, {\cal \bvec{B}}^2$ is directly proportional to the probability 
density function for a photon
\footnote{If we have a beam of $n$ equal photons each of them with 
energy $\epsilon$ (given by the Planck relation), since $\frac{1}{2} 
\, ({\cal \bvec{E}}^2 \, + \, {\cal \bvec{B}}^2)$ is the energy density 
of the electromagnetic field, then $\frac{1}{n \epsilon} \, \frac{1}{2} 
\, ({\cal \bvec{E}}^2 \, + \, {\cal \bvec{B}}^2) \, dS \, dt$ gives 
the probabilty that each photon has to be detected in the area $dS$ in 
the time $dt$. The generalization to photons of different energies 
(i.e. of different frequencies) is obtained with the aid of the 
superposition principle.}.
In terms of $\bvec{\psi}$, the Maxwell equations in vacuum then write 
\cite{Recami}
\bea
\left. \right. & \left. \right. & \bvec{\nabla} \cdot \bvec{\psi} \; = 
\; 0 \label{12} \\
\left. \right. & \left. \right. & \frac{\partial \bvec{\psi}}{\partial 
t} \; = \; i \, \bvec{\nabla} \times \bvec{\psi}
\label{13}
\eea
By making use of the correspondence principle
\bea
E & \rightarrow & i \, \frac{\partial}{\partial t} \label{14} \\
\bvec{p} & \rightarrow & - i \, \bvec{\nabla}
\label{15}
\eea
these equations effectively can be cast in a Dirac-like form
\be
\left( E \; - \; \al \cdot \bvec{p} \right) \, \bvec{\psi} \; = \; 0
\label{16}
\ee
with the transversality condition
\be
\bvec{p} \cdot \bvec{\psi} \; = \; 0
\label{17}
\ee
where the 3x3 hermitian matrices $( \alpha_i )_{lm} \, = \, i \, 
\epsilon_{i l m}$
\be
\alpha^1 \; = \;
\left( \ba{ccc}
0 & 0 & 0 \\
0 & 0 & i \\
0 & -i & 0 
\ea \right) \;\;\;\;\;\;\;\;\;\;
\alpha^2 \; = \;
\left( \ba{ccc}
0 & 0 & -i \\
0 & 0 & 0 \\
i & 0 & 0 
\ea \right) \;\;\;\;\;\;\;\;\;\;
\alpha^3 \; = \;
\left( \ba{ccc}
0 & i & 0 \\
-i & 0 & 0 \\
0 & 0 & 0 
\ea \right) \label{18}
\ee
satisfying
\be
\left[ \alpha_i \; , \; \alpha_j \right] \; = \; - i \, \epsilon_{i j k} 
\, \alpha_k \label{19} \\
\ee
have been introduced.

Note that the probabilistic interpretation is indeed possible given 
the ``continuity equation'' (Poynting theorem)
\be
\frac{\partial \rho}{\partial t} \; + \; \bvec{\nabla} \cdot \bvec{j} 
\; = \; 0
\label{110}
\ee
where
\be
\rho \; = \; \frac{1}{2} \, \bvec{\psi}^{\ast} \cdot \bvec{\psi}
\;\;\;\;\;\;\;\;\;\;\;\;\;\;\;\;
\bvec{j} \; = \; - \, \frac{1}{2} \, \psi^{\ast} \, \al \, \psi
\label{111}
\ee
are respectively the energy and momentum density of the 
electromagnetic field.

The fascination of Majorana formulation of electrodynamics lies mainly 
in the fact that it deals with directly observable quantities, such as 
the electric and the magnetic fields, without the occurence of 
electromagnetic potentials. On the other side, we know that Maxwell 
equations are Lorentz invariant, but this is not manifest in the cited 
formulation. Actually, the lack of a covariant Majorana formulation is 
due to the fact that in the known covariant formulation of 
electrodynamics it is of fundamental importance the introduction of 
the vector 4-potential $A_{\mu}$.

The main goal of this paper is to show how to obtain a genuine 
covariant Majorana formulation, without the use of the 4-potential. In 
the next section we develope the appropriate formalism to this end, 
while in section 3 the covariant Majorana-Maxwell equations are 
derived and their properties are pointed out. In section 4 we 
construct the Majorana hamiltonian and deal with spin and (intrinsic) 
boost matrices, and with the photon helicity. Finally, the conclusions.

\section{Covariant Kinematics of the Electromagnetic Field.}

Covariant electromagnetism is described by the field strength tensors 
$\fbmn$ and its dual $\ftbmn \, = \, \frac{1}{2} \, \epsilon_{\mu \nu 
\alpha \beta} \, F^{\alpha \beta}$. The basic property of these tensors 
is their antisymmetric behaviour under the exchange of their indices 
$\mu$ and $\nu$; this can be expressed saying that for any 4-vector 
$\ubm$ the relations
\be
\ubm \, \ubn \, \famn \; = \; 0 
\;\;\;\;\;\;\;\;\;\;\;\;\;\;\;\;
\ubm \, \ubn \, \ftamn \; = \; 0
\label{21}
\ee
must hold. Without loss of generality, we can assume the 4-vector 
$\ubm$ to be unitary, so that
\be
\ubm \, \uam \; = \; 1
\label{22}
\ee
Let us now introduce the following ``auxiliary fields" 
($\ubm$-dependent)
\be
\eam \; = \; \ubn \, \famn
\;\;\;\;\;\;\;\;\;\;\;\;\;\;\;\;
\bam \; = \; \ubn \, \ftamn
\label{23}
\ee
in terms of which the property (\ref{21}) writes
\be
\ubm \, \eam \; = \; 0
\;\;\;\;\;\;\;\;\;\;\;\;\;\;\;\;
\ubm \, \bam \; = \; 0
\label{24}
\ee
i.e. the 4-vectors $\ebm$, $\bbm$ are both orthogonal to $\ubm$. We 
are now able to express the tensors $\fbmn$ and $\ftbmn$ by means of 
the auxiliary fields in the following way:
\be
\fbmn \; = \; \ubm \, \ebn \; - \; \ubn \, \ebm \; + \; \epsilon_{ \mu 
\nu \alpha \beta} \, B^{\alpha} \, u^{\beta}
\label{25}
\ee
\be
\ftbmn \; = \; \ubm \, \bbn \; - \; \ubn \, \bbm \; - \; \epsilon_{ \mu 
\nu \alpha \beta} \, E^{\alpha} \, u^{\beta}
\label{26}
\ee
We stress that the fields $\ebm$, $\bbm$ depends on $\ubm$ in such a 
manner that $\fbmn$ and $\ftbmn$ are independent on $\ubm$ itself. We 
see, in fact, that the auxiliary fields $\ebm$, $\bbm$ are nothing 
that covariant linear $\ubm$-combinations of the physical electric and 
magnetic field
\be
{\cal E}^i \; = \; F^{0 i}
\;\;\;\;\;\;\;\;\;\;\;\;\;\;\;\;
{\cal B}^i \; = \; \widetilde{F}^{0 i}
\label{27}
\ee
Namely, using (\ref{22}), (\ref{24}), (\ref{25}), (\ref{26}), the 
relations between the auxiliary fields and the physical fields are the 
following:
\be
E_0 \; = \; \bvec{u} \cdot \bvec{\cal E}
\;\;\;\;\;\;\;\;\;\;\;\;\;\;\;\;\;
\bvec{E} \; = \; u_0 \, \bvec{\cal E} \; + \; \bvec{u} \times 
\bvec{\cal B}
\label{28}
\ee
\be
B_0 \; = \; \bvec{u} \cdot \bvec{\cal B}
\;\;\;\;\;\;\;\;\;\;\;\;\;\;\;\;\;
\bvec{B} \; = \; u_0 \, \bvec{\cal B} \; - \; \bvec{u} \times 
\bvec{\cal E}
\label{29}
\ee
For a given $\ubm$, the fields $\ebm$, $\bbm$ are univocally 
determined by the electric and magnetic fields. Note that in the 
``isotropic case'', in which
\be
\ubm \; = \; (1, \bvec{0})
\label{210}
\ee
we obtain $\eam \, = \, (0, \bvec{\cal E})$, $\bam \, = \, (0, 
\bvec{\cal B})$.

It is interesting to observe the analogy between the relations 
(\ref{28}), (\ref{29}) and those for the transformations of the 
electric and magnetic field from an inertial reference frame to 
another moving with relative velocity $\bvec{v}$ with $|\bvec{v}| \ll 1$ 
\cite{Landau}:
\be
\bvec{\cal E}^{\prime} \; \simeq \; \bvec{\cal E} \; + \; \bvec{v} 
\times \bvec{\cal B}
\;\;\;\;\;\;\;\;\;\;\;\;\;\;\;\;
\bvec{\cal B}^{\prime} \; \simeq \; \bvec{\cal B} \; - \; \bvec{v} 
\times \bvec{\cal E}
\label{211}
\ee
This suggestive analogy allows to interpret the 3-vector part of 
the auxiliary fields as the Lorentz transformed electric and magnetic 
fields and $\bvec{u}$ as a velocity.

\section{Covariant Majorana form of Maxwell equations.}

Armed with the formalism developed in the previous section, we are now 
able to write Maxwell equations according to Majorana's point of view. 
Here we are interested in illustrating the method, so that we will 
consider only non-interacting photons; Maxwell equations in vacuum are 
then
\be
\dbm \, \famn \; = \; 0 
\;\;\;\;\;\;\;\;\;\;\;\;\;\;\;\;
\dbm \, \ftamn \; = \; 0
\label{31}
\ee
Before continuing, let us note that equations (\ref{31}) imply
\be
\dbm \, \eam \; = \; 0 
\;\;\;\;\;\;\;\;\;\;\;\;\;\;\;\;
\dbm \, \bbm \; = \; 0
\label{32}
\ee
Now, assuming $\ubm$ independent on space-time coordinates, when we 
substitute (\ref{25}), (\ref{26}) in (\ref{31}) we have
\be
\uam \, \dbm \, \ean \; + \; \epsilon^{\mu \nu \alpha \beta} \, \dbm 
\, B_{\alpha} \, u_{\beta} \; = \; 0
\label{33}
\ee
\be
\uam \, \dbm \, \ban \; - \; \epsilon^{\mu \nu \alpha \beta} \, \dbm 
\, E_{\alpha} \, u_{\beta} \; = \; 0
\label{34}
\ee
Introducing the complex field
\be
\pbm \; = \; \ebm \; - \; i \, \bbm
\label{35}
\ee
satisfying
\be
\dbm \, \pam \; = \; 0
\label{36}
\ee
equations (\ref{31}) rewrite as
\be
\dbm \, \left( \uam \, \pan \; + \; i \, \epsilon^{\mu \nu \alpha 
\beta} \, \psi_{\alpha} \, u_{\beta} \right) \; = \; 0
\label{37}
\ee
Then, defining the following four 4x4 matrices
\be
\left( \gam \right)^{\alpha}_{\;\; \beta} \; = \; \uam \, g^{\alpha}_{\;\;
\beta} \; + \; i \, \epsilon^{\mu \alpha}_{\;\; \;\; \beta \gamma} \, 
u^{\gamma}
\label{38}
\ee
the equations of motion of the electromagnetic field in vacuum acquire 
the form (using (\ref{14}), (\ref{15}))
\be
\left( \gam \right)^{\alpha}_{\;\; \beta} \, p_{\mu} \, \psi^{\beta} \; 
= \; 0
\label{39}
\ee
This is just the ``Dirac equation'' for free photons.

\subsection{Properties of $\gamma$-matrices.}

The explicit form of the four $\gamma$-matrices is the following:
\bea
\gamma^0 & = &
\left( \ba{cccc}
u^0 & 0 & 0 & 0 \\
0 & u^0 & i \, u^3 & - i \, u^2 \\
0 & - i \, u^3 & u^0 & i \, u^1 \\
0 & i \, u^2 & - i \, u^1 & u^0
\ea \right) \\
\gamma^1 & = &
\left( \ba{cccc}
u^1 & 0 & - i \, u^3 & i \, u^2 \\
0 & u^1 & 0 & 0 \\
- i \, u^3 & 0 & u^1 & i \, u^0 \\
i \, u^2 & 0 & - i \, u^1 & u^1 
\ea \right) \\
\gamma^2 & = &
\left( \ba{cccc}
u^2 & i \, u^3 & 0 & - i \, u^1 \\
i \, u^3 & u^2 & 0 & - i \, u^0 \\
0 & 0 & u^2 & 0 \\
- i \, u^1 & i \, u^0 & 0 & u^2 
\ea \right) \\
\gamma^3 & = &
\left( \ba{cccc}
u^3 & - i \, u^2 & i \, u^1 & 0 \\
- i \, u^2 & u^3 & i \, u^0 & 0 \\
i \, u^1 & - i \, u^0 & u^3 & 0 \\
0 & 0 & 0 & u^3
\ea \right)
\eea
These are hermitian matrices
\be
\left( \left( \gam \right)_{\beta}^{\;\; \alpha} \right)^{\ast} \; = 
\; \left( \gam \right)^{\alpha}_{\;\; \beta}
\label{314}
\ee
satisfying the relations
\bea
\ubm \, \gam & = & 1 \label{315} \\
Tr \, \gam & = & 4 \, \uam \label{316}
\eea
The algebra of $\gamma$-matrices is defined by
\be
\left[ \gam \; , \; \gan \right] \; = \; \frac{i}{4} \, \epsilon^{\mu 
\nu \alpha \beta} \, \left( Tr \, \gamma_{\alpha} \right) \, \gamma_{\beta}
\label{317}
\ee
Explicitely, in terms of $\ubm$, their commutator can be expressed as 
follows:
\be
\left[ \gam \; , \; \gan \right]^{\alpha}_{\;\; \beta} \; = \;
u^{\alpha} \, \left( \uam \, g^{\nu}_{\;\; \beta} \; - \;  \uan \, 
g^{\mu}_{\;\; \beta} \right) \; + \; \left( g^{\alpha \mu} \, \uan \; 
- \; g^{\alpha \nu} \, \uam \right) \, u_{\beta} \; - \; \left( 
g^{\mu \alpha} \, g^{\nu}_{\;\; \beta} \; - \; g^{\nu \alpha} \, 
g^{\mu}_{\;\; \beta} \right)
\label{318}
\ee

\subsection{The ``isotropic case'' $\ubm \, = \, (1, \bvec{0})$.}

In the special case in which $\ubm \, = \, (1, \bvec{0})$ the 
$\gamma$-matrices reduce to
\bea
\gamma^0 & = &
\left( \ba{cccc}
1 & 0 & 0 & 0 \\
0 & 1 & 0 & 0 \\
0 & 0 & 1 & 0 \\
0 & 0 & 0 & 1 
\ea \right) \;\;\;\;\;\;\;\;\;\;\; 
\gamma^1 \; = \;
\left( \ba{cccc}
0 & 0 & 0 & 0 \\
0 & 0 & 0 & 0 \\
0 & 0 & 0 & i \\
0 & 0 & -i & 0 
\ea \right) \\
\gamma^2 & = &
\left( \ba{cccc}
0 & 0 & 0 & 0 \\
0 & 0 & 0 & -i \\
0 & 0 & 0 & 0 \\
0 & i & 0 & 0 
\ea \right) \;\;\;\;\;\;\;\;\;\;
\gamma^3 \; = \;
\left( \ba{cccc}
0 & 0 & 0 & 0 \\
0 & 0 & i & 0 \\
0 & -i & 0 & 0 \\
0 & 0 & 0 & 0 
\ea \right)
\eea
so that $\left( \gamma^i \right)^{l m} \, = \, i \, \epsilon^{i l m }$ 
coincide with the $\alpha$-matrices in (\ref{18}). The equations of 
motion then riproduce exactly the Majorana-Maxwell equations 
(\ref{12}), (\ref{13})
\bea
\left. \right. & \left. \right. & \bvec{p} \cdot \bvec{\psi} \; 
= \; 0 \label{321} \\
\left. \right. & \left. \right. & E \, \bvec{\psi} \; + \; i \, 
\bvec{p} \times \bvec{\psi} \; = \; 0 \label{322}
\eea

\subsection{Properties of the equations of motion.}

From (\ref{39}), (\ref{38}) we have:
\be
\left( u \cdot p \right) \, \pam \; = \; i \, \epsilon^{\mu}_{\;\; 
\alpha \beta \gamma} \, p^{\alpha} \, \psi^{\beta} \, u^{\gamma}
\label{323}
\ee
For $u \cdot p \, \neq  \, 0$ (in the limit (\ref{210})
this selects solutions with $E \, \neq \, 0$), multiplying (\ref{323}) 
by $\ubm$ and summing we obtain
\be
\ubm \, \pam \; = \; 0
\label{324}
\ee
In an analogous way, multiplying by $p_{\mu}$ we get also
\be
p_{\mu} \, \pam \; = \; 0
\label{325}
\ee
From these, we directly see that the equations of motions (\ref{39}) 
contains the orthogonality conditions (\ref{24}) and (\ref{36}).
For further applications, it is useful to deduce also the ``adjoint 
equation'' of (\ref{39}); by means of the hermiticity condition 
(\ref{314}), we then have
\be
\psi^{\ast}_{\alpha} \, 
\left( \gam \right)^{\alpha}_{\;\; \beta} \, p_{\mu} \; = \; 0
\label{326}
\ee
(where we understand that $p_{\mu} \, = \, i \, 
\stackrel{\leftarrow}{\dbm}$ acts on 
$\psi^{\ast}$). The comparison of (\ref{326}) and (\ref{39}) makes 
evident the fact that photons coincide with antiphotons \cite{Recami}.

\subsection{Probability current: the continuity equation.}

Multiplying (\ref{39}) by $\psi^{\ast}$ and (\ref{326}) by $\psi$ and 
summing, we easily obtain the equation
\be
\dbm \, J^{\mu} \; = \; 0
\label{327}
\ee
where the current $J^{\mu}$ is given by
\footnote{Note that the factor $\frac{1}{2}$ can be eliminated by a 
redefinition of the fields $\psi$ and $\psi^{\ast}$, while the minus 
sign can be absorbed in the definition of $\gamma$-matrices.}
\be
J^{\mu} \; = \; - \, \frac{1}{2} \, \psi^{\ast} \, \gam \; \psi
\label{328}
\ee
and satisfies
\be
\ubm \, J^{\mu} \; = \; - \, \frac{1}{2} \, \psi^{\ast} \cdot \psi
\label{329}
\ee
The ``probability density'' $J^0$ in (\ref{328}) is a well-defined 
positive quantity (as one can easily check), and in the limit 
(\ref{210}) eq. (\ref{327}) reduces to the Poynting theorem 
(\ref{110}).

However, in the general case (with an arbitrary $\ubm$) it is more 
correct to consider directly the energy-momentum tensor
\be
T^{\mu \nu} \; = \; - \, F^{\mu \sigma} \, F^{\nu}_{\;\; \sigma} \; + 
\; \frac{1}{4} \, g^{\mu \nu} \, F_{\alpha \beta} \, F^{\alpha \beta}
\label{330}
\ee
whose components $T^{0 \mu}$ are proportional to the probability 
current density, as explained in section 1 ($T^{00} \, = \, 
\frac{1}{2} \, ( \, {\cal E}^2 \, + \, {\cal B}^2 )$ , $T^{0 i} \, = 
\, {\cal E} \times {\cal B}$). Substituting (\ref{25}) and (\ref{38}) 
in (\ref{330}) we have the following expression for the tensor $T^{\mu 
\nu}$:
\be
T^{\mu \nu} \; = \; - \, \frac{1}{4} \left(
\psi^{\ast} \, \left\{ \gam \, , \, \gan \right\} \, \psi \; + \; 
\left( \psi^{\ast \; \mu} \, \pan \; + \; \psi^{\ast \; \nu} \, \pam 
\right) \right)
\label{331}
\ee
where $\{ ... \, , \, ... \}$ denotes the anticommutator. Using 
(\ref{318}), $T^{\mu \nu}$ can be cast in the asymmetric form
\be
T^{\mu \nu} \; = \; - \, \frac{1}{2} \, \left\{ 
\psi^{\ast} \,  \gam \, \gan \, \psi \; + \; 
\psi^{\ast \; \mu} \, \pan \right\}
\label{332}
\ee
Another useful writing of $T^{\mu \nu}$ in terms of $\ubm$ is the 
following:
\bea
T^{\mu \nu} & = & \left( \frac{1}{2} \, g^{\mu \nu} \, - \, \uam \, 
\uan \right) \, \psi^{\ast} \cdot \psi \; - \; \frac{i}{2} \, \left( 
\uam \, \epsilon^{\nu}_{\;\; \alpha \beta \gamma} \, + \, \uan \, 
\epsilon^{\mu}_{\;\; \alpha \beta \gamma} \right) \, \psi^{\ast \; 
\alpha} \, \psi^{\beta} \, u^{\gamma} \nonumber \\
& \left. \right. & - \; \frac{1}{2} \, \left( 
\psi^{\ast \; \mu} \, \pan \; + \; \psi^{\ast \; \nu} \, \pam 
\right)
\label{333}
\eea
Using, then, the equations of motion one can prove that $T^{\mu \nu}$ 
is conserved:
\be
\dbn \, T^{\mu \nu} \; = \; 0
\label{334}
\ee
The general expression for the probability current density is, 
finally, given by $T^{0 \mu}$.

The fact that we have found two conserved currents, $J^{\mu}$ and 
$T^{0 \mu}$, which coincide only in the limit (\ref{210}), induces to 
ask to ourselves why this happen. In reality, the conservation of 
$J^{\mu}$ and $T^{0 \mu}$ are two expressions of the same physical 
principle, that is the energy-momentum conservation (or, in other 
words, the probability conservation). In fact, one can easily prove 
that
\be
T^{\mu \nu} \; = \; \uam \, J^{\nu} \; + \; \uan \, J^{\mu} \; - \; 
\left( u \cdot J \right) \, g^{\mu \nu} \; + \; j^{\mu \nu}
\label{335}
\ee
where
\be
j^{\mu \nu} \; = \; - \, \frac{1}{2} \, \psi^{\ast} \, \tau^{\mu \nu} 
\; \psi
\label{336}
\ee
\be
\left( \tau^{\mu \nu} \right)^{\alpha}_{\;\; \beta} \; = \; g^{\mu 
\alpha} \, g^{\nu}_{\;\; \beta} \; + \; g^{\nu 
\alpha} \, g^{\mu}_{\;\; \beta} 
\label{337}
\ee
is a particular symmetric tensor satisfying
\be
\ubm \, j^{\mu \nu} \; = \; 0
\label{338}
\ee
The current $J^{\mu}$ is then given by
\be
J^{\mu} \; = \; \ubn \, T^{\mu \nu}
\label{339}
\ee
from which it is clear that the conservation of  $J^{\mu}$ follows 
from that of $T^{\mu \nu}$, as stated earlier.

A very peculiar role is played by the current $j^{0 \mu}$; as we shall 
see in the next section, this is a Lorentz boost term, whose presence 
is not evident in the non-covariant formulation of Majorana-Maxwell 
electromagnetism. Obviously, this ``boost current'' is not a conserved 
one,
\be
\dbm \, j^{0 \mu} \; = \; \partial^0 \left( u \cdot J \right) \; - \; 
\left( u \cdot \partial \right) \, J^0
\label{340}
\ee
and it is interesting to observe, from (\ref{338}), that it does not 
contribute to the current $J^{\mu}$.

\section{Hamiltonian form. Spin and intrinsic boost.}

From the covariant equation of motion (\ref{39}) one can deduce a 
useful Lorentz non-invariant hamiltonian formulation left-multiplying 
(\ref{39}) by $\gamma_0^{-1}$. The equations of motion can then be 
cast in the form $H \, \psi \, = \, E \, \psi$ with
\be
H \; = \; \bvec{a} \cdot \bvec{p} \;\;\;\;\;\;\;\;\;\;\;\;\;\;\;\;
\bvec{a} \; = \; \gamma_0^{-1} \, \bvec{\gamma}
\label{41}
\ee
The explicit form of the matrix $\gamma_0^{-1}$ is given by
\be
\left( \gamma_0^{-1} \right)^{\alpha}_{\;\; \beta} \; = \; u_0 \, 
g^{\alpha}_{\;\; \beta} \; - \; \left( g^{\alpha 0} \, u_{\beta} \; + 
\; u^{\alpha} \, g^{0}_{\;\; \beta} \right) \; + \; \frac{u^{\alpha} 
\, u_{\beta}}{u_0} \; + \; \frac{g^{\alpha 0} \, g^0_{\;\; \beta}}{u_0} 
\; - \; i \, \epsilon^{0 \alpha}_{\;\;\;\; \beta \gamma} \, u^{\gamma}
\label{42}
\ee
and the following relations hold true:
\bea
u_0 \, \left( \gamma_0^{-1} \; + \; \gamma_0 \right) & = & 
u_0 \, \left( \gamma_0^{\ast} \; - \; \gamma_0 \right) \; + \; 
\left( \gamma_0^{2} \; + \; 1 \right) \label{43}\\
\left( \gamma_0^{-1} \right)^{\dagger} & = & \gamma_0^{-1} 
\label{44}
\eea
the latter saying that $\gamma_0^{-1}$ is hermitian, as the other 
$\gamma$-matrices. It is then easy to see that, in the form 
(\ref{41}), the matrices $\bvec{a}$ are not hermitian, in general.

However, this is just an accident since, using (\ref{324}) and the 
$\mu \, = \, 0$ equation of (\ref{323}), we can recast the equations 
of motion in the form $H \, \psi \, = \, E \, \psi$ with, now,
\be
H \; = \; \left( \al \; + \; \ta \right) \cdot \bvec{p}
\label{45}
\ee
where the matrices
\bea
\alpha^1 & = &
\left( \ba{cccc}
0 & 0 & 0 & 0 \\
0 & 0 & 0 & 0 \\
0 & 0 & 0 & i \\
0 & 0 & -i & 0 
\ea \right) \;\;\;\;\;\;\;\;\;\;
\alpha^2 \; = \;
\left( \ba{cccc}
0 & 0 & 0 & 0 \\
0 & 0 & 0 & -i \\
0 & 0 & 0 & 0 \\
0 & i & 0 & 0 
\ea \right) \;\;\;\;\;\;\;\;\;\;
\alpha^3 \; = \;
\left( \ba{cccc}
0 & 0 & 0 & 0 \\
0 & 0 & i & 0 \\
0 & -i & 0 & 0 \\
0 & 0 & 0 & 0 
\ea \right) \nonumber \\
\tau^1 & = &
\left( \ba{cccc}
0 & 1 & 0 & 0 \\
1 & 0 & 0 & 0 \\
0 & 0 & 0 & 0 \\
0 & 0 & 0 & 0 
\ea \right) \;\;\;\;\;\;\;\;\;\;\;\;
\tau^2 \; = \;
\left( \ba{cccc}
0 & 0 & 1 & 0 \\
0 & 0 & 0 & 0 \\
1 & 0 & 0 & 0 \\
0 & 0 & 0 & 0 
\ea \right) \;\;\;\;\;\;\;\;\;\;\;\;\;
\tau^3 \; = \;
\left( \ba{cccc}
0 & 0 & 0 & 1 \\
0 & 0 & 0 & 0 \\
0 & 0 & 0 & 0 \\
1 & 0 & 0 & 0 
\ea \right) \nonumber
\eea
are involved. Note that 
\be
\left( \alpha^i \right)^{\alpha}_{\;\; \beta} \; = \; - i \, 
\epsilon^{0 i \alpha}_{\;\;\;\;\;\; \beta}
\label{48}
\ee
are the generalization to 4 dimensions of the $\al$-matrices in 
(\ref{18}), while
\be
\left( \tau^i \right)^{\alpha}_{\;\; \beta} \; = \; g^{0 \alpha} \, 
g^{i}_{\;\; \beta} \; + \; g^{i \alpha} \, g^0_{\;\; \beta}
\label{49}
\ee
coincide with the matrices $\tau^{0 i}$ defined in (\ref{337}).

It is somewhat interesting to observe that the hamiltonian in 
(\ref{45}) is explicitely independent on $\ubm$ (and holds true for 
arbitrary $\ubm$), the equations of motion 
$H \, \psi \, = \, E \, \psi$ being themselves explicitely independent 
on $\ubm$:
\bea
E \, \psi_0 & = & \bvec{p} \cdot 
\bvec{\psi} \; \label{410} \\
E \, \bvec{\psi} & = & 
\bvec{p} \, \psi_0 \; - i \, 
\bvec{p} \times \bvec{\psi} \label{411}
\eea
However, not all the solutions of (\ref{410}), (\ref{411}) are 
allowed, since in this form these equations does not contain the 
orthogonality condition (\ref{324}). This is a fundamental condition, 
expressing the antisymmetric behaviour of $\fbmn$ , $\ftbmn$; so 
together with
(\ref{410}), (\ref{411}) also the equation (\ref{324}) must be 
considered, and the allowed solutions of 
$H \, \psi \, = \, E \, \psi$ are those satisfying (\ref{324}).

Obviously, in the limit (\ref{210}) we recover the Majorana-Maxwell 
equations (\ref{321}), (\ref{322}).

Let us now turn to the matrices (\ref{48}) and (\ref{49}). These 
verifies the following commutation rules:
\bea
\left[ \alpha^i \; , \; \alpha^j \right] & = & - i \, \epsilon^{i j k} 
\, \alpha^k \label{412} \\
\left[ \alpha^i \; , \; \tau^j \right] & = & - i \, \epsilon^{i j k} 
\, \tau^k \label{413} \\
\left[ \tau^i \; , \; \tau^j \right] & = & - i \, \epsilon^{i j k} 
\, \alpha^k \label{414}
\eea
and we recognize in these the Lorentz algebra.

By requiring that the total angular momentum of the photon $\bvec{J} 
\, = \, \bvec{L} \, + \, \bvec{S}$ is conserved (i.e. $[ H \, , \, 
\bvec{J} ] \, = \, 0$) we can then identify the spin matrices with
\be
\bvec{S} \; = \; - \, \al
\label{415}
\ee
In an analogous way, we can further identify the intrinsic part of 
Lorentz boost matrices with
\be
\bvec{K}_s \; = \; i \, \ta
\label{416}
\ee
as we have already anticipated.

With the introduction of the spin matrices, we can now rewrite the 
equations of motion in the form
\be
\frac{\bvec{S} \cdot \bvec{p}}{E} \; \psi \; = \; - \, \psi \; + \; 
\frac{\bvec{\tau} \cdot \bvec{p}}{E} \; \psi
\label{417}
\ee
where in the L.H.S. we recognize the helicity operator. With a direct 
calculus, using (\ref{325}), we see that
\be
\frac{\bvec{\tau}^{\mu}_{\;\; \nu} \cdot \bvec{p}}{E} \; \pan \; = \; 
\frac{p^{\mu}}{E^2} \, \left( \bvec{p} \cdot \bvec{\psi} \right)
\label{418}
\ee
so that the boost term in the equations of motion is proportional to 
$\bvec{p} \cdot \bvec{\psi}$. {\it If} $\bvec{p} \cdot \bvec{\psi} \, 
= \, 0$ the boost term vanishes, and then {\it only} in this case the 
photon is in an helicity eigenstate (with $\lambda \, = \, \pm 1$ 
eigenvalues). But, from (\ref{325}), $\bvec{p} \cdot \bvec{\psi} \, 
= \, 0$ means $\psi_0 \, = \, 0$ and this , from (\ref{324}), implies 
that $\ubm \, = \, (1 , \bvec{0})$.

We then conclude that the physical transversality of the photon 
implies:
\begin{itemize}
\item the photon is in an helicity eigenstate (with $\lambda \, = \, 
\pm 1$ eigenvalues);
\item a Lorentz boost term in the equations of motion is forbidden.
\end{itemize}
Inversely:
\begin{itemize}
\item if the photon has also longitudinal degrees of freedom, then it 
is not in an helicity eigenstate (as happens for the Dirac equation 
too);
\item longitudinal degrees of freedom ({\it if present} \cite{Evans}) 
are described by a Lorentz boost term in the equations of motion.
\end{itemize}

\section{Conclusions.}

In this paper we have given a version of the Majorana formulation of 
Maxwell electromagnetism which is explicitely invariant under the 
Lorentz group. The main advantage of the Majorana formulation is that 
it does not require the use of electromagnetic potentials, thus 
avoiding non-physical degrees of freedom. This has been here achieved 
also in the explicit Lorentz invariant version with the introduction 
of ``auxiliary fielda'', linear (covariant) combinations of the 
electric and magnetic fields built using the basic antisymmetric 
properties of the electromagnetic field strenghts $\fbmn$ and 
$\ftbmn$.

As also in the non-covariant version \cite{Majo, Recami}, the Majorana 
formulation of electromagnetism is strongly based on the fact that 
there are 3 spatial and 1 temporal dimensions, since only in 3+1 
dimensions the dual tensor of $\fbmn$ is a two indices tensor as 
$\fbmn$.

Here we have shown not only that Maxwell equations can be written as a 
Dirac-like equation for the photon, but also that the interpretative 
structure of the Dirac equation can be maintained for the photon as 
well. In fact, interpreting $\psi^{\ast}$ as the adjoint field of 
$\psi$ (see eq. (\ref{326})), the probability current for the photon 
has the same form as in the Dirac case (cfr. eq. (\ref{328})), 
provided the new $\gamma$-matrices are defined as in (\ref{38}).

The ``transversality'' property of the photon (\ref{325}) (together 
with (\ref{324})) is contained in the Dirac-like equation (\ref{39}) 
and it is thus not necessary a further constraint, as in the 
non-covariant formulation (see eq. (\ref{17})).

Finally, the spin matrices, but also the (intrinsic) boost ones, have 
been derived in section 4 using the hamiltonian form of the Dirac-like 
equation (\ref{39}), and the helicity properties of the photon has 
been recovered as well.

\vspace{1truecm}
\noindent
{\Large \bf Acknowledgements}\\
\noindent
This paper takes its origin by a fruitful mailing correspondence with 
Prof. E. Recami: the author is indebted with him.

\end{document}